# Computation of Trusted Short Weierstrass Elliptic Curves for Cryptography

*Kunal Abhishek*[1], *E. George Dharma Prakash Raj*[2]

[1]*Society for Electronic Transactions and Security, Chennai, India*
[2]*Bharathidasan University, Tiruchirappalli, India*
*E-mails*: *kunalabh@gmail.com* *georgeprakashraj@yahoo.com*

***Abstract*****:** *Short Weierstrass elliptic curves with underlying hard Elliptic Curve Discrete Logarithm Problem* (*ECDLP*) *are widely used in cryptographic applications. A notion of security called Elliptic Curve Cryptography* (*ECC*) *security is also suggested in literature to safeguard the elliptic curve cryptosystems from their implementation flaws. In this paper, a new security notion called the "trusted security" is introduced for computational method of elliptic curves for cryptography. We propose three additional "trusted security acceptance criteria" which need to be met by the elliptic curves aimed for cryptography. Further, two cryptographically secure elliptic curves over 256 bit and 384 bit prime fields are demonstrated which are secure from ECDLP, ECC as well as trust perspectives. The proposed elliptic curves are successfully subjected to thorough security analysis and performance evaluation with respect to key generation and signing/verification and hence, proven for their cryptographic suitability and great feasibility for acceptance by the community.*

***Keywords*****:** *Short Weierstrass elliptic curves, prime field, cryptography, ECDLP Security, ECC Security, Trusted Security.*

## 1. Introduction

Short Weierstrass elliptic curves are considered to be as secure for cryptography as the underlying hardness of their Elliptic Curve Discrete Logarithm Problem, i.e., (ECDLP) which is defined as finding a scalar $k$ knowing any two points $P$ and $Q$ on elliptic curve $\mathbb{E}$ holding the relation $Q = kP$. This is known as the ECDLP security of the selected elliptic curve when used for cryptography [1]. The most efficient publicly known method to solve ECDLP or break the ECDLP security is the Pollard's rho algorithm which takes approximately $0.886\sqrt{n}$ point additions where $n$ is the base point order [1-2]. One must select an elliptic curve which is ECDLP secure for cryptographic applications. Another notion of security for selecting suitable elliptic curves for cryptography is known as elliptic curve cryptography security, i.e., ECC security in short, the term coined by Bernstein and Lange [1] which ensures prevention from any information leakage from the implementation flaws of the elliptic curve.

Most of the popular standards today such as National Institute of Standards and Technology (NIST) [3], Brainpool [4], Standards for Efficient Cryptography 2 (SEC2) [5], IEEE P1363 [6], etc., recommended those elliptic curves which are ECDLP secure and attain some sort of ECC security (for only some standard curves [1]). It is worthwhile to note that an ECC based cryptosystem can be compromised by either compromising the ECDLP security or the ECC security. All the present day standards have recommended Short Weierstrass elliptic curves keeping either or both of these security notions into consideration. This paper introduces a critical security notion which we call as "trusted security" of elliptic curves which ensures that the selected elliptic curve is free from any manipulation from its computation perspective and can be trusted for use in cryptographic applications. The trusted security notion of computation of elliptic curves minimizes the risks involved in generation of safe curve parameters deterministically where they are vulnerable to (intentionally) non-disclosed attacks with (intentionally) non-disclosed properties of the curve parameters. In such cases, the ECDLP can be solvable by using very efficient sub-exponential or polynomial time algorithm using non-guessable high computing power.

The key contributions of this paper are as follows:

1. Introduction of a new security notion called as "trusted security acceptance criteria" as an important security evaluation criterion along with the ECDLP security and ECC security criteria for computation of Short Weierstrass elliptic curves aimed for cryptography.

2. Evaluation of standard Short Weierstrass elliptic curves from trust perspective.

3. Argument that trust in generation method of elliptic curves can be achieved only through computation of the curve parameters randomly without considering any of their pre-studied values such as $a = -3$ or $p$ as Mersenne primes, etc. The randomly selected elliptic curve parameters can be derived using any good quality user trusted Random Number Generator (RNG) along with competitive curve performance.

4. Demonstration of two new elliptic curves called as Kunal-George 256 bit first random elliptic curve (KG256r1) and Kunal-George 384 bit first random elliptic curve (KG384r1) defined over 256 bit and 384 bit prime field sizes respectively for cryptography which are secure from ECDLP security, ECC security as well as trusted security perspectives.

5. Evaluation of the proposed elliptic curves KG256r1 and KG384r1 with respect to cryptographic key pair generation, signing and verification from performance perspective.

Organization of the paper is as follows.

Section 2 deals with the background and problem statements of the presented work. Section 3 introduces the proposed "trusted security acceptance criteria" for cryptographically safe elliptic curve computation. Section 4 evaluates standard Short Weierstrass elliptic curves from trusted security acceptance criteria perspective.

Section 5 describes the generation procedure including the proposed trusted security acceptance criteria to derive new elliptic curves KG256r1 and KG384r1 for evaluation and demonstration. Section 5 also holds the discussion on importance of trusted security acceptance criteria of elliptic curves to minimize the risk of manipulating the curve parameters intended for cryptographic purposes. Section 6 presents demonstration of the proposed trusted Short Weierstrass elliptic curves for cryptography. Section 7 gives the security analysis of the proposed elliptic curves. Section 8 discusses results obtained in the presented work and demonstration of the performance metrics of the proposed elliptic curves. Finally, Section 9 concludes the paper and gives future directions.

## 2. Background and problem statements

An elliptic curve in Short Weierstrass form consists of three parameters: a prime number $p$ which is the order of the underlying field over which the elliptic curve is defined and two field coefficients $a$ and $b$. The formal definition of a Short Weierstrass elliptic curve and its twisted curve are as follows:

**Definition 1 [7].** A Short Weierstrass elliptic curve $\mathbb{E}(\mathbb{F}_p)$ of prime field order $p$ is the set of all solutions $(x, y)$ to the equation

(1) $$\mathbb{E}: y^2 = x^3 + ax + b,$$

where $a$, $b$ are the coefficients in $\mathbb{F}_p$ with field characteristic greater than 3. The elliptic curve $\mathbb{E}$ also includes a special point $\mathbb{O}$ called the point at infinity. $\mathbb{E}$ has non-singularity condition, i.e., its discriminant $\triangle_{\mathbb{E}} = 4a^3 + 27b^2 \neq 0$.

The field order $p$ determines the security level offered by the elliptic curve. Hence, it is important to select $p$ as big as possible. Generally, $p \geq 256$ bits in size gives accepted security level while $p$ of 256 bit length is considered as widely accepted prime field size of the elliptic curve for interoperability purposes.

**Definition 2 [8].** If $\mathbb{E}: y^2 = x^3 + ax + b$ be an elliptic curve with $a, b \in \mathbb{F}_p$ the twist of $\mathbb{E}$ by $c \in \mathbb{F}_p$ is defined as

(2) $$\mathbb{E}': y^2 = x^3 + ac^2x + bc^3.$$

It is important to select those elliptic curves which are cryptographically secure and trusted for constructing cryptographic systems. Transport Layer Security (TLS), Secure SHell (SSH) and Internet Protocol Security (IPSec) [9], Public Key Infrastructure (PKI) [10], etc., are some of the popular applications which require safe elliptic curves in their cryptosystem design. Most of such commercial applications use standard elliptic curves over prime field of 256 bit sizes for sufficient security and interoperability purposes. However, Bernstein et al. [2] have recently pointed out some mechanisms such that a new elliptic curve can be proposed to sabotage public standards. They demonstrated convincing methods by which they were able to implant vulnerability in the elliptic curves known as BADA55 curves by utilizing the gain of many bits of freedom [2] which satisfies the public standards and can be put forward for standardization to fool the users. This essentially proves that an attacker can exploit unknown (known to him) vulnerability to sabotage existing public standards and justify his selection of elliptic curve parameters citing performance gain and his own way of getting randomness, i.e., verifiably random, etc., which is used in the generation of the vulnerable curve parameters. Bernstein et al. [2] comprehensively demonstrated how a wrong or non-trustable elliptic curve can be derived using the procedure led by the public standards and their recommended public criteria. They showed that plausible variations in the Brainpool curve generation procedure and Microsoft curve generation procedure respectively can be used to sabotage public standard. Further, the Agence Nationale de la Securite des Systemes d'Information (ANSSI) standard recommended FRP256V1 elliptic curve which has low twist security of order $2^{79}$ which means that there are $2^{79}$ elliptic curve additions required to mount the twist attack to get user's secret key [2]. Also, there is no reasonably sufficient documentation available for this curve. Furthermore, Bernstein et. al. [2] demonstrated computation of the BADA55-R-256 curve which meets the public security criteria for ECDLP security and ECC security but still is a manipulated curve. Finally, we understand that computation of an elliptic curve can be manipulated by any deterministic method of computation of the curve parameters and variety of reasons can be cited with selection of the curve parameters adhering to some public standard of proposer's convenience.

Summarizing, the problems pertained with the trust consists of one or more issue(s) from the following:

- No sufficient explanation on the RNG used for seed or randomness generation.
- Intentional variation in standard elliptic curve generation procedure recommended by the curve proposing agencies by themselves.
- Intentional hiding of information about the curve parameters even providing detailed documentations on curve generation process of standard elliptic curves.
- Sabotaged standards.
- Root problem of the lack of trust is the deterministic approach adopted by all the agencies in standardizing their proposed elliptic curves.

With the above prevalent issues, an obvious question arises that "because you can explain, does not mean that you will explain everything". We answer this question by introducing a set of three important security evaluation criteria called "trusted security acceptance criteria" for computation of suitable elliptic curves for cryptography which can be additionally invoked along with the ECDLP security and ECC security criteria to mitigate the trust issues in curve generation process to a great extent.

## 3. Trusted security acceptance criteria for elliptic curves for cryptography

Standard elliptic curves followed deterministic approach in computation of their coefficients and primes. Most of them used pre-studied values whose credibility and trustworthiness are doubted [2, 11-13] due to origination of the curve parameters and lack of proof for the randomness used in the curve generation process such as use of computationally convenient primes like powers of two, etc. Hence, there is a need to introduce additional security acceptability criteria to invoke trust in the computation of elliptic curve parameters for use and in standardization. In this paper, a set of three new security evaluation criteria of cryptographically safe elliptic curve called the "trusted security acceptance criteria" for elliptic curve used for cryptography is introduced which is as follows:

a. T1: User trusted Random Number Generator (RNG) to provide (pseudo)randomness.

A RNG should be selected preferably by its user for assuring that user is fully aware of the technicality of the RNG and hence he/she trusts it completely. Apart from the trust aspect, the RNG should adhere to the following properties as indicated by Koc [14] and Schneier [15]:

- The bitstream generated by a PseudoRandom Number Generator (PRNG) or Cryptographically Secure PRNG (CSPRNG) should be statistically sound, i.e., it has a large period.
- The bitstream generated should be unpredictable, i.e., the RNG should be forward secure as well as backward secure.

The curve parameters should be chosen randomly in a trustworthy way to avoid any uneasy explanation about the generation of the curve constants and hence, the requirement of user trusted and strong RNG is critical in trust building.

b. T2: No pre-studied values of the curve coefficients and prime.

The well-known constants are accepted by everyone without hesitation but their non-exposed property may be used for construction of vulnerable elliptic curves. BADA55-VPR-224 is such an example which used $\cos(1)$ constant [2]. The elliptic curve coefficients $a$, $b$ must not use any pre-studied values to avoid the scope of manipulation. Moreover, the prime field order $p$ can only have special structure if it is randomly selected with suitable size (normally $\geq 224$) bits for fast reduction on the elliptic curve.

c. T3: Reproducibility of new elliptic curves of nearly the same cryptographic strength and suitability using the same method and apparatus.

One must get new elliptic curves of nearly the same cryptographic strength using the same method and apparatus. We consider Pollard's rho values of the elliptic curves and their respective twisted curves as the measurement of their cryptographic strengths which is the number of elliptic curve point additions to solve the ECDLP. Generally, $0.886\sqrt{n}$ elliptic curve point additions are required to break the ECDLP where $n$ is the order of the base point [1-2].

## 4. Evaluation of standard elliptic curves from trust perspective

Standard Short Weierstrass elliptic curves claimed to have followed rigorous ECDLP security validations and sometime ECC security validations together to arrive at the curve parameters for recommendation. They claimed that they used seeds which were randomly generated and some of them adhered to verifiably random way of obtaining the curve parameters. Table 1 evaluates standard elliptic curves from trust perspectives for use in cryptography.

Table 1. Evaluation of the standard Short Weierstrass elliptic curves from trust perspective

| Elliptic curve | Trusted Security (T1, T2, T3) | Remarks |
|---|---|---|
| NIST P224r1 | None | Deterministic approach with pre-studied coefficients and prime [3] |
| NIST P256r1 | None | Deterministic approach with pre-studied coefficients and prime [3] |
| NIST P384r1 | None | Deterministic approach with pre-studied coefficients and prime [3] |
| secp224r1 | None | Special structure of prime $p$ (Mersenne prime) and insufficient documentation [5] |
| secp256r1 | None | Special structure of prime $p$ (Mersenne prime) and insufficient documentation [5] |
| secp384r1 | None | Special structure of prime $p$ (Mersenne prime) and insufficient documentation [5] |
| secp521r1 | None | Special structure of prime $p$ (Mersenne prime) and insufficient documentation [5] |
| ANSSI FRP256v1 curve | None | Pre-studied value of coefficient $a$ and insufficient documentation [2, 16] |
| Brainpool | T2 | None of the Brainpool curves are generated by their own stipulated procedure [2, 4] |
| NUMS curves | None | Deterministic approach with pre-studied coefficients and prime [2, 17] |

It is imperative to note from Table 1 that, there is an ardent need for new elliptic curves which are cryptographically secure as well as trusted. Following section will focus on the generation details of trusted Short Weierstrass elliptic curves to be used for cryptography.

## 5. Cryptographically secure elliptic curve generation using the proposed trusted security acceptance criteria

Short Weierstrass elliptic curves have a unique property that it can only exhibit prime order [18] in order to get maximum security of ECDLP without compromising any bit of security [19]. However, elliptic curves of cryptographic interest must get validated against their ECDLP security, ECC security as well as trusted security. It is now observed from previous sections that random approach of computing safe elliptic curves is the only way to achieve all of these three security notions. A standard procedure is shown as the flow chart in Fig. 1 for a bird’s eye view of generation of the trusted Short Weierstrass elliptic curves intended for cryptography.

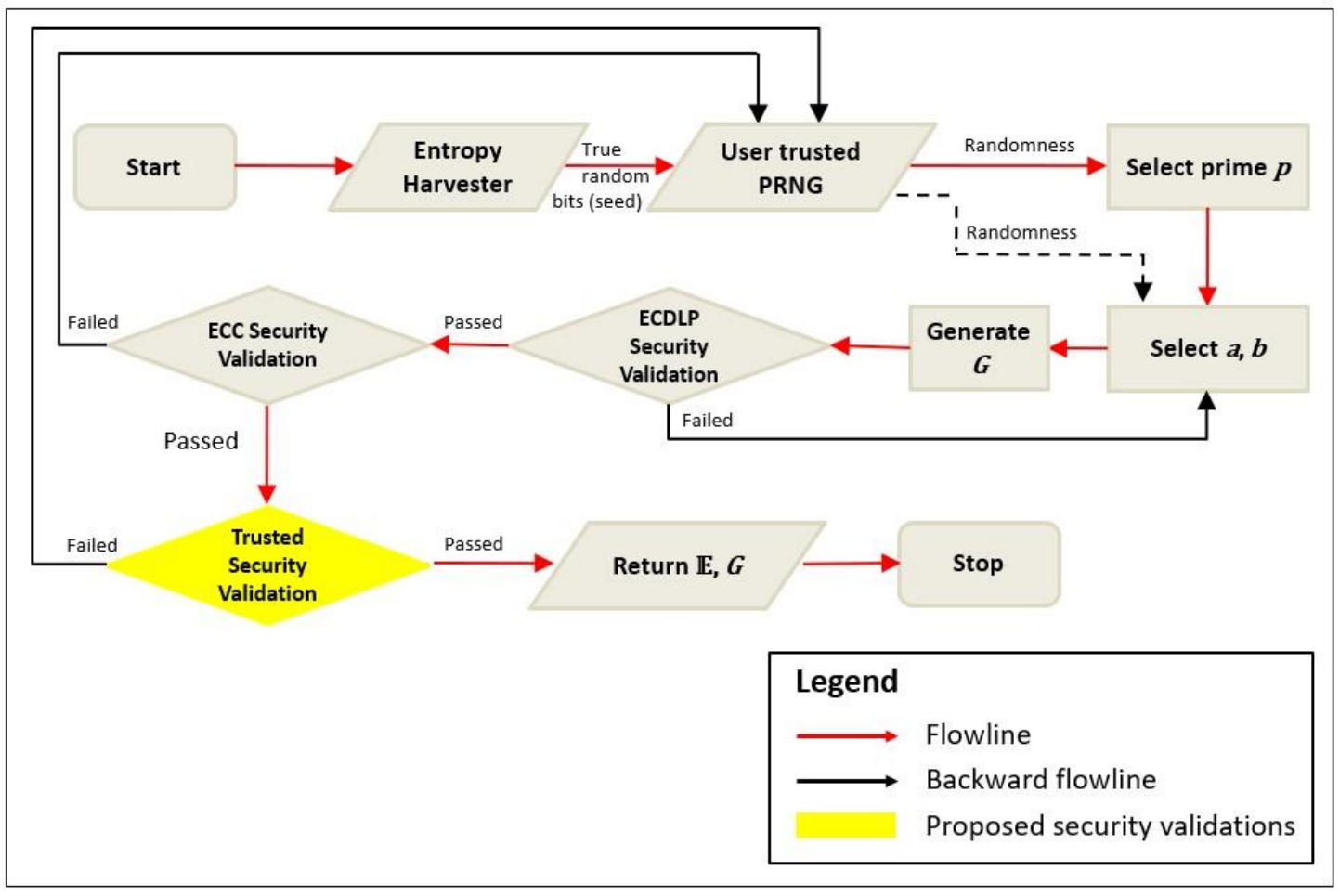

Fig. 1. Flow chart of generation of cryptographically secure and trusted Short Weierstrass elliptic curve

An entropy harvester which is used to obtain sufficient number of true random bits from various physical noise sources like device randomness, disk randomness, Human Interface Device (HID) (key board, mouse, etc.), interrupt randomness, etc., is used to seed a user trusted (means user is aware of the technicality of the RNG and associated security risks completely) PRNG/CSPRNG as depicted in Fig. 1. The user trusted PRNG supplies desired number of (pseudo)random bits to generate suitable $p$, $a$ and $b$. An elliptic curve $\mathbb{E}$ is constructed over prime field $p$ (which is fixed in our case, but one can choose other way also to generate suitable elliptic curves by fixing the curve order $N$ randomly, etc.) with coefficients $a$ and $b$. Now $\mathbb{E}$ is subjected to ECDLP security validation failing which it will regenerate the coefficients $a$ and $b$ until it gets suitable curve coefficients for $\mathbb{E}$ to be ECDLP secure. A base point $G$ is also selected randomly over elliptic curve $\mathbb{E}$ and gets verified for its prime order for acceptability. Once $\mathbb{E}$ is validated for ECDLP security, it is subjected to security validation from ECC security perspective which expects $\mathbb{E}$ to have its twist $\mathbb{E}$' also to be as secure as $\mathbb{E}$ is. In case of the fact that ECC security validation does not pass, one needs to regenerate the prime $p$ and subsequently coefficients $a$ and $b$ to get ECDLP security and ECC security successfully validated. Finally, the ECDLP secure and ECC secure $\mathbb{E}$ is verified with the proposed trusted security acceptance criteria (indicated in yellow decision box in Fig. 1) failing which the process is re-initiated with deriving prime $p$ and coefficients $a$ and $b$ as fresh until one gets an acceptable $\mathbb{E}$. Lastly, $\mathbb{E}$ and $G$ are returned as the output. The elliptic curve generation procedure is detailed in Algorithm 1.

## 5.1. Assumptions

Following assumptions were made considered while computing the curve parameters using Algorithm 1:

i. User trusted cryptographically strong RNG is available to provide randomness required in computation of secure elliptic curve.

ii. Sufficient entropy is available in the system. Generally, more than 2000 bits of entropy is expected to be available with the system to seed the RNG sufficiently to uninterruptedly generate elliptic curves up to over 384 bit prime field sizes. Also, the operating system is not used for the first time after installation as sufficient entropy will not be available with the machine.

iii. Compilers, CPU Processors, SAGE and other participating modules in the curve parameter generation are trusted.

5.2. Standard elliptic curve generation procedure including trusted security acceptance criteria

Algorithm 1 shows the standard procedure along with the proposed trusted security acceptance criteria as discussed in Fig. 1 with detailed security validations of elliptic curve from ECDLP security, ECC security and trusted security perspectives.

**Algorithm 1.** Generation of trusted cryptographically safe Short Weierstrass elliptic curve

*Input:* Prime field size (l) in bits and randomness from user trusted RNG

*Output:* Trusted cryptographically safe elliptic curve $\mathbb{E}$ over prime field $p$ with base point $G_{x,y}$

**Step 1.** Input prime field size l in bits

**Step 2.** Obtain seed $S$ as true random bits of desired length from entropy harvester

**Step 3.** Set seed $S$ for user trusted RNG

**Step 4.** Select randomly prime $p$ such that $p \equiv 3 \bmod 4$ // for fast arithmetic on $\mathbb{E}$

**Step 5.** Choose randomly the coefficient $a$ of $\mathbb{E}$

**Step 6.** Choose randomly the coefficient $b$ of $\mathbb{E}$

**Step 7.** Construct the elliptic curve $\mathbb{E}$ with curve parameters $p$, $a$ and $b$

**Step 8.** Enforce ECDLP security validation:

**Step 8.1.** If discriminant $\triangle_{\mathbb{E}} = 4a^3 + 27b^2 \neq 0$ // $\mathbb{E}$ must be non-singular

**Step 8.2.** Else go to Step 5

**Step 8.3.** If curve order $N$ is prime

**Step 8.4.** Else go to Step 5

**Step 8.5.** If $\mathbb{E}$ is non-anomalous case // $N \neq p$

**Step 8.6.** Else go to Step 5

**Step 8.7.** If $\mathbb{E}$ is not supersingular curve

**Step 8.8.** Else go to Step 5

**Step 8.9.** Generate randomly the base point $G_{x,y}$ on $\mathbb{E}$

**Step 8.10.** Validate if base point order $n$ is prime

**Step 8.11.** Else go to Step 8.9

**Step 8.12.** If cofactor is 1

**Step 8.13.** Else go to Step 5

**Step 8.14.** If Pollard's rho value < $2^{100}$

**Step 8.15.** Else go to Step 5

**Step 8.16.** If embedding degree $k \geq (N-1)/100$ // guarantees intractability of DLP

**Step 8.17.** Else go to Step 5

**Step 9.** Enforce ECC security validation: (If $\mathbb{E}$ is twist secure, i.e., all validations in Step 8 applied to $\mathbb{E}'$)

**Step 9.1.** If twist discriminant $\triangle_{\mathbb{E}'}$ of $\mathbb{E} = 4a^3 + 27b^2 \neq 0$

**Step 9.2.** Else go to Step 4

**Step 9.3.** If order of $\mathbb{E}'$, $N$ is prime

**Step 9.4.** Else go to Step 4

**Step 9.5.** If $\mathbb{E}'$ is non-anomalous case

**Step 9.6.** Else go to Step 4

**Step 9.7.** If $\mathbb{E}'$ is not supersingular curve

**Step 9.8.** Else go to Step 4

**Step 9.9.** Generate randomly the base point $G'_{x,y}$ on $\mathbb{E}'$

**Step 9.10.** Validate if base point order $n'$ is prime

**Step 9.11.** Else go to Step 9.9

**Step 9.12.** If cofactor of $\mathbb{E}'$ is 1

**Step 9.13.** Else go to Step 4

**Step 9.14.** If Pollard's rho value of $\mathbb{E}' < 2^{100}$

**Step 9.15.** Else go to Step 4

**Step 9.16.** If embedding degree $k' \geq (N' - 1)/100$

**Step 9.17.** Else go to Step 4

**Step 10.** Enforce trusted security validation

**Step 10.1.** Validate if RNG is trusted // Proposed validation criterion T1

**Step 10.2.** Else go to Step 2

**Step 10.3.** Validate if coefficients $a$ and $b$ have no pre-studied value // Proposed validation criterion T2

**Step 10.4.** Else go to Step 2

**Step 10.5.** Validate if elliptic curves with similar cryptographic strength can be generated with the same method and apparatus // Proposed validation criterion T3

**Step 10.6.** Else go to Step 2

**Step 11.** Return $\mathbb{E}$: $\{p, a, b\}$ and $G_{x,y}$

Algorithm 1 takes elliptic curve field size (l) in bits as the input in Step 1. A seed $S$ is extracted from the entropy harvester in Step 2. In our case, we used /dev/random as the PRNG which takes true random bits through a Hardware based RNG (HRNG) that extracts entropy directly. We used /dev/random PRNG available with Linux Fedora kernel Version 4.13.9 for obtaining randomness in desired bit lengths. The HRNG uses various noise sources like input randomness, device randomness, disk randomness, HID (key board, mouse, etc.), interrupt randomness to provide random bits as the seed $S$ to /dev/random in Step 3. $S$ is used to initialize /dev/random to provide randomness to the curve generation process as and when required. As the curve generation program needs a user trusted secure RNG, we leave it to the user to select his/her trusted RNG for fulfilling the randomness requirements. Here our focus is to recommend users to use their own trusted RNGs to avoid any possible manipulation in curve computation and we demonstrate how a trusted Short Weierstrass elliptic curve can be generated for cryptography. In Step 4, the prime $p$ of user desired l bit length is randomly selected and subsequently, checked that it should hold the form of $p \equiv 3 \bmod 4$ for fast reduction, i.e., fast elliptic curve arithmetic on $\mathbb{E}$. It is noted that $p$ is first chosen randomly and then verified for this form to avoid any pre-studied value. The curve coefficients $a$ and $b$ are then chosen randomly in Step 5 and Step 6 respectively using different seeds, i.e., $a$ and $b$ have independent initializations. Now, an elliptic curve $\mathbb{E}$ is constructed with $p$, $a$ and $b$ in Step 7.

The ECDLP security validations are enforced in Step 8 which includes non-singularity in Step 8.1, prime curve order in Step 8.3, non-anomalous property in Step 8.5, non-supersingularity in Step 8.7, random selection of base point in Step 8.8 with prime base point order in Step 8.9, small cofactor as 1 in Step 8.11, high Pollard's rho in Step 8.14 and high embedding degree in Step 8.16 respectively. Non-singularity of elliptic curve confirms that curve is smooth and indeed an elliptic curve [20-22]. Prime order elliptic curve with order $N$ is resistant to Pohlig-Hellman attack when $N \geq 2^{160}$ [23]. Non-anomalous case of elliptic curve, i.e., when curve order $N \neq p$, confirms that curve is resistant to pairing based attacks [23]. Non-supersingularity of elliptic curve prevents the ECDLP from the Menezes, Okamoto and Vanstone (MOV) reduction attack with sub-exponential complexity which takes place when the conditions that $p$ divides trace $t$ or/and $t^2 = 0, p, 2p, 3p$ or $4p$ are met [24-25]. The cofactor value determines the cryptographic security and gives maximum security when selected as 1 [23, 25]. The Pollard's rho value of elliptic curve determines the number of elliptic curve point additions to find a collision. This check is very important as a parallelized Pollard-rho on $r$ processors can solve ECDLP in $(\sqrt{\pi n})/\sqrt{2r}$ steps [23, 26]. The embedding degree of elliptic curve $k \geq 20$ is considered sufficient to guarantee intractability of the discrete logarithm problem in the extension field [7].

The ECC security validations are enforced in Step 9 of Algorithm 1 in which it looks for the twist of the selected elliptic curve to be secure against all the ECDLP security validations as described above. The twist security of elliptic curve prevents from any implementation flaws or information leakage about the user's secret key [1].

The trusted security validations are carried out in Step 10 to ensure the method of generation of elliptic curve is trusted in terms of the randomness used in the curve generation process and the curve parameters are drawn randomly. It also ensures that the procedure described in Algorithm 1 can be used to obtain Short Weierstrass elliptic curves of nearly the same cryptographic strength each time on its execution.

Finally, a trusted and secure elliptic curve $\mathbb{E}$: $\{p, a, b\}$ and base point $G_{x,y}$ is returned in Step 11.

## 6. Demonstration of trusted Short Weierstrass elliptic curves

We used Algorithm 1 to derive two trusted Short Weierstrass elliptic curves KG256r1 and KG384r1 defined over 256 bit and 384 bit respectively for demonstration. The details of the proposed KG256r1 and KG384r1 is shown in Table 2 and Table 3, respectively. These elliptic curves have undergone security analysis in Section 7 to ensure that the elliptic curves generated using Algorithm 1 have nearly the same cryptographic strength in terms of Pollard's rho complexity and other criteria like big discriminant, embedding degree, trace, etc., while being compliant with the three security notions, i.e., ECDLP security, ECC security and trusted security.

Table 2. The proposed KG256r1 elliptic curve

| KG256r1 | |
|---|---|
| $p$ | 105659876450476807015340827963890761976980048986351025435035631207814085532543 |
| $a$ | 57780130698115176583488499171344771088898507337873238590400955371129685138826 |
| $b$ | 102451950841073747949316796495896937960702115486975363798323596797327090813462 |
| $N$ | 105659876450476807015340827963890761976544313325663770762399235394744121359871 |
| $G$ | (5385166333114646497810998074612415985821986371151485954586014078688791960064, 88440166531789946723126083546750633179866039092883764784041611065547926159080) |
| $h$ | 1 (smallest cofactor) |

Table 3. The proposed KG384r1 elliptic curve

| KG384r1 | |
|---|---|
| $p$ | 308504936566801493400799664217561138887972017059009663818402880868888024111765 87972020735012523469267564505420764051 |
| $a$ | 268937684885793435941799884521325825414071666675195106719690165313905189264848 5257788827989185822359193013251735562 |
| $b$ | 282679914441081045194064979674986566053141057529253438397674572433074909758239 5451638354661270280127278365677483939 |
| $N$ | 308504936566801493400799664217561138887972017059009663818414387546839003900776 17323565554872996073979103765917522731 |
| $G$ | (263821674697227290786867915392591910846306526222054061903021467945234141274511 83423914120811487055055064792875345576, 202628051316606152195895866462280785015451818341996421511941020893449272958898 57293563989127020260020122002404045204) |
| $h$ | 1 (smallest cofactor) |

**Resources used.** The curve generation programme was written in Python language using Python Version 2 and Python Version 3.6 compilers and ran on a desktop server having 2*Intel® Xeon® E5-2620v4 processor with 32 CPU cores and 2.1 GHz clock frequency and 128 GB DDR4 memory. The desktop server was equipped with Linux Fedora operating system (kernel Version 4.13.9) and SAGE Version 8.1 was used for number theory arithmetic support for the curve generation program.

## 7. Security analysis of the proposed KG256r1 and KG384r1 elliptic curves

### 7.1. Analysis of the ECDLP and ECC security of the proposed KG256r1 and KG384r1 elliptic curves

We used SafeCurves verification script [1] to verify ECDLP security and ECC security of the elliptic curve parameters. Algorithm 2 describes the SafeCurves verification script which was used to verify the KG256r1 and KG384r1 elliptic curves against its ECDLP and ECC security.

**Algorithm 2.** Verification of the proposed elliptic curve parameters for cryptographic security

*Input:* Elliptic curve parameters $p, a, b, N, G_{x,y}$

*Output:* Safe/Weak Elliptic Curve

**Step 1.** Verify if shape of elliptic curve is Short Weierstrass
**Step 2.** Else return “Not Short Weierstrass elliptic curve”
**Step 3.** Verify if $p$ is prime
**Step 4.** Else return “Weak elliptic curve”
**Step 5.** Verify if discriminant $< -2^{100}$
**Step 6.** Else return “Weak elliptic curve”
**Step 7.** Verify if base point order is prime
**Step 8.** Else return “Weak elliptic curve”
**Step 9.** Verify if GCD (Curve order, base point order)=1
**Step 10.** Else return “Weak elliptic curve”
**Step 11.** Verify if base point is on curve
**Step 12.** Else return “Incorrect base point”
**Step 13.** Verify if co-factor is 1 or 2 or 4
**Step 14.** Else return “Weak elliptic curve”
**Step 15.** Verify if $p+1-t$ is a multiple of base point order $n$
**Step 16.** Else return “Weak elliptic curve”
**Step 17.** Verify if embedding degree of curve $\geq (N-1)/100$
**Step 18.** Else return “Weak elliptic curve”
**Step 19.** Verify if elliptic curve is MOV safe
**Step 20.** Else return “Weak elliptic curve”
**Step 21.** Verify if base point order of twist $!= p$
**Step 22.** Else return “Weak elliptic curve”
**Step 23.** Verify if twist equation is elliptic
**Step 24.** Else return “Weak elliptic curve”
**Step 25.** Verify if twist shape is Short Weierstrass
**Step 26.** Else return “Weak elliptic curve”
**Step 27.** Verify co-factor of twist is 1 or 2 or 4
**Step 28.** Else return “Weak elliptic curve”
**Step 29.** Verify if GCD (base point order of twist, $p$) $= 1$
**Step 30.** Else return “Weak elliptic curve”
**Step 31.** Verify if Pollard’s rho value of elliptic curve $\geq 2^{100}$
**Step 32.** Else return “Weak elliptic curve”
**Step 33.** Verify if rigidity is True
**Step 34.** Else return “Weak elliptic curve”
**Step 35.** Verify if twist rho value $\geq 2^{100}$
**Step 36.** Else return “Weak elliptic curve”
**Step 37.** Verify if Joint Rho $\geq 2^{100}$
**Step 38.** Else return “Weak elliptic curve”
**Step 39.** Otherwise, return “Cryptographically safe elliptic curve”

It is obvious that ECDLP security is a critical security requirement for qualifying any elliptic curve for cryptography. However, SafeCurves [1] proposed ECC security as another security notion for evaluating elliptic curves to ensure that the ECC implementations do not reveal or leak information about user's secret key. For Short Weierstrass elliptic curves, a twist secure elliptic curve can prevent ECC implementation flaws [1]. The elliptic curve $\mathbb{E}$ is twist secure if its twist $\mathbb{E}'$ is secure which means that all the ECDLP security validations are also successfully compliant by $\mathbb{E}'$ [1].

Both KG256r1 and KG384r1 elliptic curves qualified all the ECDLP and ECC security verifications executed in Algorithm 2. The field orders $p$ and curve orders $N$ of both elliptic curves were verified deterministically for being a prime number using Pocklington's theorem [1]. We avoided any special structure of prime or pre-studied value to prevent from any vulnerability. For example, NIST P-224 prime, i.e., $p = 2^{224} - 2^{96} + 1$ was used by BADA55-VPR-224 and standard ANSSI prime

0xF1FD178C0B3AD58F10126DE8CE42435B3961ADBCABC8CA6DE8FCF353D86E9C03

was used by BADA55-R-256 curve, respectively, to demonstrate vulnerable curves to the community [2]. Moreover, the discriminants, embedding degrees, cofactor values and Pollard's rho values of both curves and their respective twist curves were verified successfully possessing more than their expected threshold values. These curves were also verified to confirm that they are not a case of anomalous and supersingular ones as discussed in Section 5.2 and thus, they are suitable for cryptography. Table 4 and Table 5 shows these values possessed by both KG256r1 and KG384r1 elliptic curves.

## 7.2. Analysis of trusted security of KG256r1 and KG384r1 elliptic curves

### 7.2.1. Validation of Trusted Security Criteria T1

We trust and used /dev/random PRNG for curve generation procedure due to the fact that it has faced a lot of successful cryptanalysis [27-29] and sustained long with the Linux kernel since 1994 [28]. Moreover, the latest versions (Version 4.8 or later) of /dev/random have overcome [30] the criticism of having possible entropy attacks [2]. We used Linux Fedora kernel Version 4.13.9 and selected /dev/random as the PRNG (sometimes /dev/random is referred as True Random Number Generator (TRNG) because it has the direct interface with the HRNG). We are actually making a point here that choose your trusted RNG and own the risk associated with your selection.

### 7.2.2. Validation of Trusted Security Criteria T2

To validate the T2 criterion, no pre-studied values of the curve coefficients $a$ and $b$ are used as they have been chosen randomly and independently. The prime numbers $p$ in both proposed curves KG256r1 and KG384r1 are selected randomly first and then chosen with a form of $p \equiv 3 \bmod 4$ for performance tuning and there is no evidence of these primes $p$ and coefficients $a$ and $b$ reported in past as the pre-studied ones.

Table 4. Verification result of the ECDLP security of the proposed elliptic curves

| Elliptic curve $\mathbb{E}$ | Offered security level | Rho complexity ($\rho$-value) | Embedding degree ($k$) | Trace ($t$) | Discriminant ($D$) | Curve order ($N$) | Co-factor ($h$) | Non-anomalous? | Non-supersingular? |
|---|---|---|---|---|---|---|---|---|---|
| KG256r1 | 128 | 127.8 | 1056598764504768070153408279638907619765443133256637707623992353947441213598 7 | 4357356606872546726363958130699641726 73 | –232773939807348890850436587536448542043730245608155363125035103075438982165243 | 105659876450476807015340827963890761976544313325663770762399235394744121359871 ($N > 2^{256}$) | 1 | Yes | Yes |
| KG384r1 | 192 | 191.6 | 30850493656680149340079966421756113888797201705900966381841438754683900390077617323565554872996073979103765917522 73 | –1150667795097978901029351544819860472604711539260496758679 | –1220779382520449530033023314772621110455402982992783892893127974464429030246303129345660706643594391150137565212 31163 | 308504936566801493400799664217561138887972017059009663818414387546839003900776173235655548729960739791037659175227 31 ($N > 2^{384}$) | 1 | Yes | Yes |

Table 5. Verification result of the ECC security of the proposed elliptic curves

| Twist of elliptic curve $\mathbb{E}'$ | Offered security level in bits | Rho complexity ($\rho'$-value) | Embedding degree ($k'$) | Curve order ($N'$) | Co-factor ($h'$) | Non-anomalous? | Non-supersingular? |
|---|---|---|---|---|---|---|---|
| KG256r1 | 128 | 127.8 | 440249485210320029230586783182878174905899102695992833781966779253683540 4384 | 10565987645047680701534082796389076197741578464703828010767202702088404970 5217 ($N' > 2^{256}$) | 1 | Yes | Yes |
| KG384r1 | 192 | 191.6 | 308504936566801493400799664217561138887972017059009663818391374190937044322755586204759151520508645560252449240053 72 | 308504936566801493400799664217561138887972017059009663818391374190937044322755586204759151520508645560252449240053 73 ($N' > 2^{384}$) | 1 | Yes | Yes |

### 7.2.3. Validation of Trusted Security Criteria T3

To validate the T3 criterion, we conducted an experiment by taking three trials of executing Algorithm 1 under the same operational environment with same method and apparatus to retrieve three independent elliptic curves of the same prime lengths. Subsequently, we examined if they exhibit nearly the same cryptographic strength measured in terms of Pollard's rho value for the curves and their respective twists as discussed in Section 5.2. Table 6 shows the results obtained from this experiment which proves the successful validation of T3 criterion by the proposed KG256r1 and KG384r1 elliptic curves.

Table 6. Validation of Trusted Security criteria of three new elliptic curves: T3

| Trial number | Elliptic curve parameters | Pollard’s rho value/Twist rho value |
|---|---|---|
| 1 | *p*: 87052253706622316800662279631344302713612816742118516 445715106163825624186987 | Rho: $2^{127.6}$<br>Twist rho: $2^{127.6}$ |
| | *a*: 17461513680488110202189680065467433355982187313809984308530183605390654503146 | |
| | *b*: 4742364534479307087696244304071666435175166931536995811081067226406616322940 | |
| | $G_{x,y}$: (34562444864263447792289881666782368199808912751831663386444135083641970670103, 44973717098200324632781286735408077067884851416905001940895476727480258436423) | |
| 2 | *p*: 83857931886285555818472058950522827195247211639379970952195176566538052148959 | Rho: $2^{127.6}$<br>Twist rho: $2^{127.6}$ |
| | *a*: 15222031410359054028041793088708374885174581007053672026416069700422500171995 | |
| | *b*: 75723663712830868158926603330488486312788754915163584116380630010872983931491 | |
| | $G_{x,y}$: (79991145613299850861660922601873046504314421039422310330231620709939495217575, 74048930300595054686355764380599730714484651315014966555673263252180995491420) | |
| 3 | *p*: 115455173683647336766695198555386616062185957400074700902465398650769617153383 | Rho: $2^{127.8}$<br>Twist rho: $2^{127.8}$ |
| | *a*: 89247089594531861167221907824679361896477781827771349654639873760799894221702 | |
| | *b*: 47456080838438598020722203116343582455579601993324094611207713288744264819618 | |
| | $G_{x,y}$: (87380972861908942926601892812209714038534482432156502027178728221855540030831, 109010224703610275807776999662587399010415605756892207650540783549332069147687) | |

# 8. Results and discussion

The proposed elliptic curves KG256r1 and KG384r1 are compared with other similar standard Short Weierstrass elliptic curves like NIST, SEC2, Brainpool, FRP256v1 and NUMS curves from ECDLP security, ECC security and trusted security perspectives in this section.

## 8.1. Comparison of the proposed KG256r1 and KG384r1 elliptic curves with standard elliptic curves from ECDLP and ECC security perspectives

It is imperative to note from Table 7 that none of the standard elliptic curves have passed all the SafeCurves verification criteria [1] of ECDLP security and ECC security. However, Brainpool recommended elliptic curves have deviated in their own stipulated procedure of generation [2] and hence cannot be trusted easily. Also, their verifiably random generation method is under question as such thing can be intentionally implanted to manipulate the standard as demonstrated by Bernstein et. al. through BADA55 curves [2].

Table 7. Comparison of ECDLP Security and ECC Security of the standard elliptic curves and the proposed elliptic curves [1]

| Verification criterion | Details | Supported by the curve |
|---|---|---|
| SafeField | Prime of the forms 1 mod 4 and 3 mod 4 | A, B, C, D1, KG256r1, KG384r1 |
| safeEquation | Elliptic curve over prime field possessing either Short Weierstrass or Montgomery or Edward equation | A, B, C, D1, KG256r1, KG384r1 |
| safeBase | Possessing prime order of base point | A, B, C, D1, KG256r1, KG384r1 |
| safeRho | Rho value must be $\geq 2^{100}$ | A, B, C, D1, KG256r1, KG384r1 |
| safeTransfer | Resistant to Smart-ASS attack (additive transfer) and MOV attack (multiplicative transfer) | A, B, C, D1, KG256r1, KG384r1 |
| safeDiscriminant | Absolute value of complex-multiplication field discriminant $\lvert D \rvert > 2^{100}$ | A, B, D1, KG256r1, KG384r1 |
| safeRigid | Allows only fully rigid and somewhat rigid curves | B, C, KG256r1, KG384r1 |
| safeTwist | Above security requirements for twist of the curve as well | C, KG256r1, KG384r1 |
| safeCurve | Elliptic curve is safe if all the above criteria are met | KG256r1, KG384r1 |

Note: A = NIST recommended elliptic curves, B = Brainpool recommended elliptic curves, C = SEC2 elliptic curves, D1 = ANSSI recommended elliptic curve FRP256v1.

## 8.2. Comparison of cryptographic security of the proposed KG256r1 and KG384r1 with standard elliptic curves

Table 8. Comparative security evaluation of the proposed elliptic curves with the standard elliptic curves

| Elliptic curve | ECDLP security | ECC security | Trusted security (T1, T2, T3) | Remarks |
|---|---|---|---|---|
| NIST P224r1 | Yes | No | No | No RNG description. Pre-studied value of coefficient $a$ and special structure of prime $p$ in Mersenne form. Weak twist security [3] |
| NIST P256r1 | Yes | No | No | No RNG description. Pre-studied value of coefficient $a$ and special structure of prime $p$ in Mersenne form. Weak twist security [3] |
| NIST P384r1 | Yes | No | No | No RNG description. Pre-studied value of coefficient $a$ and special structure of prime $p$ in Mersenne form. Weak twist security [3] |
| SEC2 prime curves | Yes | No | No | Special structure of prime $p$ (Mersenne prime) and insufficient documentation [5] |
| Brainpool curves | Yes | No | No | None of the Brainpool curves are generated by their own stipulated procedure [2] |
| ANSSI FRP256v1 curve | Yes | No | No | Pre-studied value of coefficient $a$ and insufficient documentation [2] |
| NUMS curve | Yes | No | No | Deterministic approach with pre-studied coefficients and prime [18] |
| KG256r1 | Yes | Yes | Yes | Randomly generated curve parameters with no pre-studied value. User trusted RNG to minimize the risk of manipulation |
| KG384r1 | Yes | Yes | Yes | Randomly generated curve parameters with no pre-studied value. User trusted RNG to minimize the risk of manipulation |

The proposed elliptic curves KG256r1 and KG384r1 are compared with standard Short Weierstrass elliptic curves from overall security of ECDLP, ECC and trust perspectives in Table 8.

We observe from Table 8 that only the proposed KG256r1 and KG384r1 elliptic curves are secure from ECDLP, ECC and trust perspectives and standard elliptic curves have met the ECDLP security validations only.

### 8.3. Performance of the proposed elliptic curves

The proposed KG256r1 and KG384r1 elliptic curves demonstrated with cryptographic operations like key pair generation, signing and verification on desktop machine having x86_64 with Intel(R) Core(TM) i5-10400 CPU with 2.90GHz processor, 16GB DDR4 memory using GNU/Linux version 5.4.0-58-generic and Python Version 3.8.5 software library. Table 9 shows the performance metrics of the proposed elliptic curves in cryptographic implementations such as key pair generation, signing and verification. The values indicated are the average of 10,000 trials' outcomes.

Table 9. Performance of the proposed elliptic curves in cryptographic implementations

| Proposed elliptic curve | Key pair generation | | Signing | | Verification | |
|---|---|---|---|---|---|---|
| | Time elapsed (in s) | Number of CPU clock cycles used | Time elapsed (in s) | Number of CPU clock cycles used | Time elapsed (in s) | Number of CPU clock cycles used |
| KG256r1 | 0.021468 | 62,260,026 | 0.0215207 | 62,410,198 | 0.0426380 | 123,650,476 |
| KG384r1 | 0.035866 | 104,012,382 | 0.035838 | 103,931,139 | 0.106852 | 309,871,025 |

## 9. Conclusion and future directions

Three new trusted security acceptance criteria T1, T2, T3 are proposed to compute cryptographically safe elliptic curves over the prime fields. These trusted security acceptance criteria or simply, the trusted security criteria are invoked along with the ECDLP security and ECC security in order to minimize the scope of manipulation in the curve parameters due to some (intentionally) non-disclosed property or methods exhibited by their proposers and sabotaged standards. We also discussed in detail that only the randomly drawn curve parameters will have the trust factor where a user trusted strong RNG plays a critical role. The choice of selection of RNG is left with the users who will own the risks associated with his chosen RNG to generate the seed and randomness for curve parameters generation requirements. We also computed two new elliptic curves called KG256r1 and KG384r1 after validating them through the newly proposed trusted security acceptance criteria along with the ECDLP and ECC security validations. Furthermore, we experimentally proved that if elliptic curves are generated keeping these three security notions into consideration then they would have nearly the same cryptographic strength in terms of Pollard's rho complexity and trustworthiness or suitability. Hence, it is inferred that one must verify trusted security acceptance criteria for randomly generated elliptic curves in addition to ECDLP and ECC security validations for secure implementation of elliptic curve based cryptosystems.

The proposed argument of trusted security and demonstrated KG256r1 and KG384r1 elliptic curves gives the feasibility of future standardization of such randomly generated elliptic curves for trusted cryptographic implementations.

***Acknowledgements*: **The authors would like to thank Society for Electronic Transactions and Security (SETS) and Dr. N. Sarat Chandra Babu, Executive Director, SETS for giving the opportunity to conduct the research and write this article. Authors would also like to sincerely thank Dr. Ananda Mohan P. V. and Dr. Reshmi T. R. for their valuable suggestions and Mr. Santhosh Kumar T. for his help in experimentation.

## References

1. Bernstein, D. J., T. Lange. SafeCurves: Choosing Safe Curves for Elliptic-Curve Cryptography. Accessed 31 January 2021.
**https://safecurves.cr.yp.to**
2. Bernstein, D. J., T. Chou, C. Chuengsatiansup, A. Hülsing, E. Lambooij, T. Lange, R. Niederhagen, C. van Vredendaal. How to Manipulate Curve Standards: A White Paper for the Black Hat. – In: International Conference on Research in Security Standardisation, Springer, Cham, 15 December 2015, pp. 109-139.
**http://bada55.cr.yp.to**
3. National Institute for Standards and Technology. FIPS PUB 186-2: Digital Signature Standard, 2000. Accessed 31 January 2021.
**http://csrc.nist.gov/publications/fips/archive/fips186-2/fips186-2.pdf.**
4. Lochter, M., J. Merkle. RFC 5639: Elliptic Curve Cryptography (ECC) Brainpool Standard Curves and Curve Generation. 2010. Accessed 31 January 2021.
**https://tools. ietf.org/html/rfc5639**
5. Certicom Research. SEC 2: Recommended Elliptic Curve Domain Parameters. Version 1.0. 2000. Accessed 31 January 2021.
**http://www.secg.org/SEC2-Ver-1.0.pdf**
6. Institute of Electrical and Electronics Engineers. IEEE 1363-2000: Standard Specifications for Public Key Cryptography, 2000. Accessed 31 January 2021.
**http: //grouper.ieee.org/groups/1363/P1363/draft.html**
7. Koblitz, A. H., N. Koblitz, A. Menezes. Elliptic Curve Cryptography: The Serpentine Course of a Paradigm Shift. – Journal of Number Theory, Vol. **131**, 2011, No 5, pp. 781-814.
8. Savaş, E., T. A. Schmidt, C. K. Koç. Generating Elliptic Curves of Prime Order. – In: International Workshop on Cryptographic Hardware and Embedded Systems, Berlin, Heidelberg, Springer, May 2001, pp. 142-158.
9. Valenta, L., N. Sullivan, A. Sanso, N. Heninger. In Search of CurveSwap: Measuring Elliptic Curve Implementations in the Wild. – In: 2018 IEEE European Symposium on Security and Privacy (EuroS&P), IEEE, April 2018, pp. 384-398.
10. Caelli, W. J., E. P. Dawson, S. A. Rea. PKI, Elliptic Curve Cryptography, and Digital Signatures. – Computers & Security, Vol. **18**, 1999, No 1, pp. 47-66.
11. Shumow, D., N. Ferguson. On the Possibility of a Back Door in the NIST sp800-90 Dual Ec Prng. – In: Proc. Crypto, Vol. 7, 2007.
12. Hales, T. C. The NSA Back Door to NIST. – Notices of the AMS, Vol. **61**, 2013, No 2, pp. 190-192.
13. Bernstein, D. J., T. Lange. Security Dangers of the NIST Curves. – In: Invited Talk, International State of the Art Cryptography Workshop, Athens, Greece, 2013.

14. K o c, C. K. About Cryptographic Engineering. – In: Cryptographic Engineering, Boston, MA, Springer, 2009, pp. 1-4.
15. S c h n e i e r, B. Applied Cryptography: Protocols, Algorithms, and Source Code in C. Second Edition. John Wiley & Sons, 2007.
16. Agence nationale de la s´ecurit´e des syst`emes d'information. Publication d'un param´etrage de courbe elliptique visant des applications de passeport ´electronique et de l'administration ´electronique fran¸caise, 2011.
**https://tinyurl.com/nhog26h**
17. B o s, J. W., C. C o s t e l l o, P. L o n g a, M. N a e h r i g. Selecting Elliptic Curves for Cryptography: An Efficiency and Security Analysis. – Journal of Cryptographic Engineering, 2015, pp. 1-28.
18. C o s t e l l o, C., P. L o n g a, M. N a e h r i g. A Brief Discussion on Selecting New Elliptic Curves. Microsoft Research. Microsoft. 8 Jun 2015.
19. B o s, J. W., C. C o s t e l l o, P. L o n g a, M. N a e h r i g. Selecting Elliptic Curves for Cryptography: An Efficiency and Security Analysis – Journal of Cryptographic Engineering, Vol. **6**, November 2016, No 4, pp. 259-286.
20. C h e n g, Q. Hard Problems of Algebraic Geometry Codes. – IEEE Transactions on Information Theory, Vol. **54**, 2008, No 1, pp. 402-406.
21. M c I v o r, C. J., M. M c L o o n e, J. V. M c C a n n y. Hardware Elliptic Curve Cryptographic Processor Over rmGF(p). – IEEE Transactions on Circuits and Systems, Vol. **53**, 2006, No 9, pp. 1946-1957.
22. S c h o o f, R. Elliptic Curves Over Finite Fields and the Computation of Square Roots mod p. – Mathematics of Computation, Vol. **44**, 1985, No 170, pp. 483-494.
23. H a n k e r s o n, D., A. M e n e z e s, S. V a n s t o n e. Guide to Elliptic Curve Cryptography. Springer, 2003. 332 p. (web). ISBN: 0-387-95273-X.
24. M e n e z e s, A. J., T. O k a m o t o, S. A. V a n s t o n e. Reducing Elliptic Curve Logarithms to Logarithms in a Finite Field. – IEEE Transactions on Information Theory, Vol. **39**, 1993, No 5, pp. 1639-1646.
25. S m a r t, N. P. The Discrete Logarithm Problem on Elliptic Curves of Trace One. – Journal of Cryptology, Vol. **12**, 1999, No 3, pp.193-196.
26. V a n O o r s c h o t, P., M. W i e n e r. Parallel Collision Search with Cryptanalytic Applications. – Journal of Cryptology, Vol. **12**, 1999, pp. 1-28.
27. V i e g a, J. Practical Random Number Generation in Software. – In: Proc. of 19th Annual Computer Security Applications Conference, 2003, IEEE, 8 December 2003, pp. 129-140.
28. D o d i s, Y., D. P o i n t c h e v a l, S. R u h a u l t, D. V e r g n i a u d, D. W i c h s. Security Analysis of Pseudo-Random Number Generators with Input: /Dev/Random is not Robust. – In: Proc. of 2013 ACM SIGSAC Conference on Computer & Communications Security, 4 November 2013, pp. 647-658.
29. G u t t e r m a n, Z., B. P i n k a s, T. R e i n m a n. Analysis of the Linux Random Number Generator. – In: 2006 IEEE Symposium on Security and Privacy (S&P'06), IEEE, 21 May 2006, pp. 15-32.
30. **https://www.2uo.de/myths-about-urandom/**